\documentclass[twocolumn,prl,superscriptaddress,floatfix,showpacs]{revtex4-1}
\usepackage[utf8]{inputenc}
\usepackage{amsmath}
\usepackage{amssymb}
\usepackage{graphicx}

\usepackage{graphics}
\usepackage{epsfig}

\newcommand{\be}{\begin{equation}} \newcommand{\ee}{\end{equation}}
\newcommand{\bea}{\begin{eqnarray}} \newcommand{\eea}{\end{eqnarray}}

\begin{document}

\title{Critical phenomena on $k$-booklets}

\author{Peter Grassberger}

\affiliation{JSC, FZ J\"ulich, D-52425 J\"ulich, Germany}

\date{\today}
\begin{abstract}
We define a `$k$-booklet' to be a set of $k$ semi-infinite planes with $-\infty < x < \infty$ and 
$y \geq 0$, glued together at the edges (the `spine') $y=0$.
On such booklets we study three critical phenomena: Self-avoiding random walks, the Ising model, and 
percolation. For $k=2$ a booklet is equivalent to a single infinite lattice, for $k=1$ to a 
semi-infinite lattice. In both these cases the systems show standard critical phenomena. This is not 
so for $k\geq 3$. Self avoiding walks starting at $y=0$ show a first order transition at a shifted
critical point, with no power-behaved scaling laws. The Ising model and percolation show hybrid transitions,
i.e. the scaling laws of the standard models coexist with discontinuities of the order parameter
at $y\approx 0$, and the critical points are not shifted. In case of the Ising model ergodicity is 
already broken at $T=T_c$, and not only for $T<T_c$ as in the standard geometry. In all three models 
correlations (as measured by walk and cluster shapes) are highly anisotropic for small $y$.
\end{abstract}
\maketitle


Critical phenomena are usually considered either on regular lattices or on random graphs. What happens
if these are replaced by geometries that are regular lattices {\it nearly} everywhere, but are atypical 
at subdomains of measure zero? 

The best known such cases are semi-infinite lattices with $d$ bulk dimensions and with one $d-1$-dimensional
surface. The Ising model on a simple cubic lattice restricted to $z=0$ is maybe the most studied model
of that type. If the surface bonds are not much stronger than the bulk bonds (``ordinary transition"), 
the magnetization slightly below $T_c$ is weaker at the surface than in the bulk and the surface order 
parameter exponent is larger than the bulk exponent. $\beta_1 > \beta$. In the other extreme of very 
strong surface bonds (``extra-ordinary transition"), the surface can already order when the bulk is still 
disordered and the scaling at $T_c$ is that of the 
``normal" transition with an applied non-zero magnetic field at the surface \cite{Diehl}.

The situation is very similar for other critical phenomena such as percolation or self-avoiding walks (SAW's).
In all these cases there is a special point where the strength of the surface bonds (or contacts in case of 
SAW's) is just sufficient to compensate the disordering effect of the absent bonds with $z < 0$. If the 
surface bonds/contacts are stronger than the special value, the surface orders already when the bulk is 
still disordered.

This scenario has to be modified for $d=2$, where the surface has $d=1$. In that case neither the 
Ising model nor percolation can order at the surface at any finite control parameter. Thus extra-ordinary 
and special transitions do not exist for them, while they do exist for SAW's \cite{Batchelor}.

But this is not yet the end of the story. Consider $k$ semi-infinite planes with $y \geq 0$, glued together at 
the edges $y=0$. If $k>2$, we can expect that random fluctuations of the order parameter at $y=0$ can reinforce 
each other. If going from $k=2$ (infinite plane) to $k=1$ (semi-infinite plane) tends to disorder the lattice,
we can envisage that going to $k>2$ will lead to increased ordering.
This is indeed true, but the details are non-trivial. 

For simplicity, we shall in the following consider only square lattices. Locally, the only difference between 
lattice sites on and off the spine is just their coordination number. While it is 4 for all sites off 
the spine, it is $2+k$ on the spine. Apart from this, the spine is treated in the algorithms used in 
this paper (which are all growth algorithms) like the rest of the system.

After this work was completed, I learned that the Ising model on $k$-booklets (called there ``multiple junctions") 
had been studied before by means of renormalization group methods \cite{Igloi:1993,Igloi:1993a}. Recently,
also directed percolation was studied on multiple junctions \cite{Igloi:2016}. Although the methods used in
\cite{Igloi:1993,Igloi:1993a} were very different, as are also the details observed in \cite{Igloi:2016}, these 
papers fully support our results.

\begin{figure}
\begin{centering}
\includegraphics[scale=0.30]{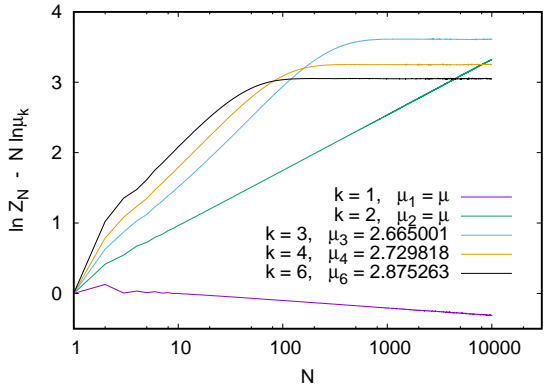}
\par\end{centering}
\caption{\label{fig1} (color online)  Log-log plot of average numbers of self-avoiding walks of length $N$ on $k$ joined 
  semi-infinite planes, divided by $\mu_k^N$. Here $\mu_1 = \mu_2 = \mu$ is the known connective constant of
  SAW's on ordinary square lattices, while $\mu_k$ for $k>2$ are connective constants fitted from these data.
  For $k\leq 2$ the $\gamma$-exponents are in agreement with $\gamma_1=61/64$ \cite{Cardy} and 
  $\gamma_2=\gamma=43/32$ \cite{Nienhuis}, while they are equal to 1 for $k>2$.}
\end{figure}

{\it Self avoiding walks}: For SAW's on the infinite square lattice, the number of walks of length $N$ scales 
as 
\be
    Z_N \sim N^{\gamma-1} \mu^N,
\ee
where the {\it connective constant} is $\mu = 2.63815853\ldots$ \cite{Jensen2003}, and the entropic exponent is 
$\gamma = 43/32$ \cite{Nienhuis}. Therefore, when plotting $Z_N/\mu^N$ against $N$ we expect a power law with 
exponent $\gamma-1 \approx 0.344$. Analogously we assume that the number of SAW's starting at $y=0$ for any $k$ 
increases with connective constant $\mu_k$, up to possible powers of $N$. In Fig.~1 we show 
$\ln Z_N -\mu_k^N$, obtained with the PERM algorithm \cite{PERM}, for $N$ up to $N=10 000$ and for 
$k=1,2,3,4$, and 6. For $k=1$ and $k=2$ we have $\mu_k=\mu$, and $\gamma$ is non-trivial. On the other hand, 
for $k>2$ the fitted values of $\mu_k$ increase with $k$, while $\gamma =1$. 

\begin{figure}
\begin{centering}
\includegraphics[scale=0.30]{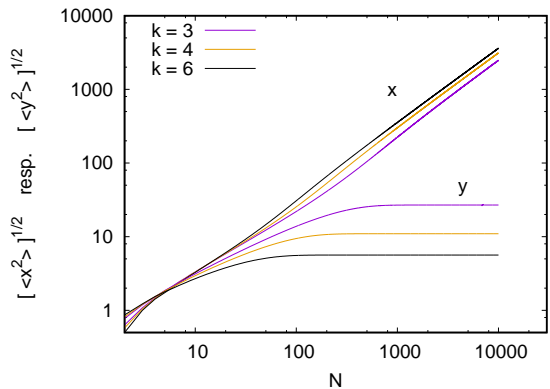}
\par\end{centering}
\caption{\label{fig2} (color online)  Log-log plot of r.m.s. end-to-end distances of self avoiding walks in 
  the directions parallel and perpendicular to the spine, for $k=3, 4,$ and 6. We do not show the data for 
  $k=1$ and 2, since they scale according to the well known Flory exponent, $[\langle x^2\rangle]^{1/2} \sim 
  [\langle y^2\rangle]^{1/2} \sim N^{3/4}$.}
\end{figure}

\begin{figure}
\begin{centering}
\includegraphics[scale=0.30]{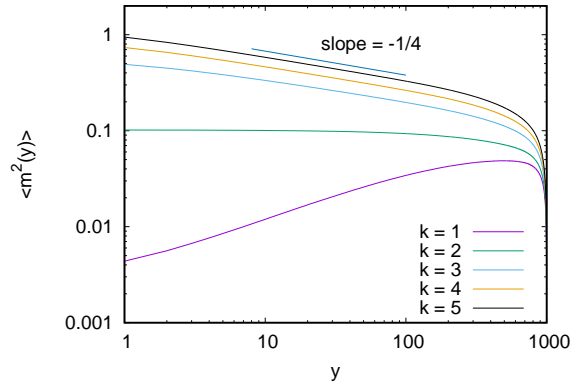}
\par\end{centering}
\caption{\label{fig3} (color online)  Log-log plot of the average squared magnetization density
  at fixed $y$, for $k$ Ising models joined at $y=0$. The temperature is the critical one for single 
  lattices.}
\end{figure}

\begin{figure}
\begin{centering}
\includegraphics[scale=0.30]{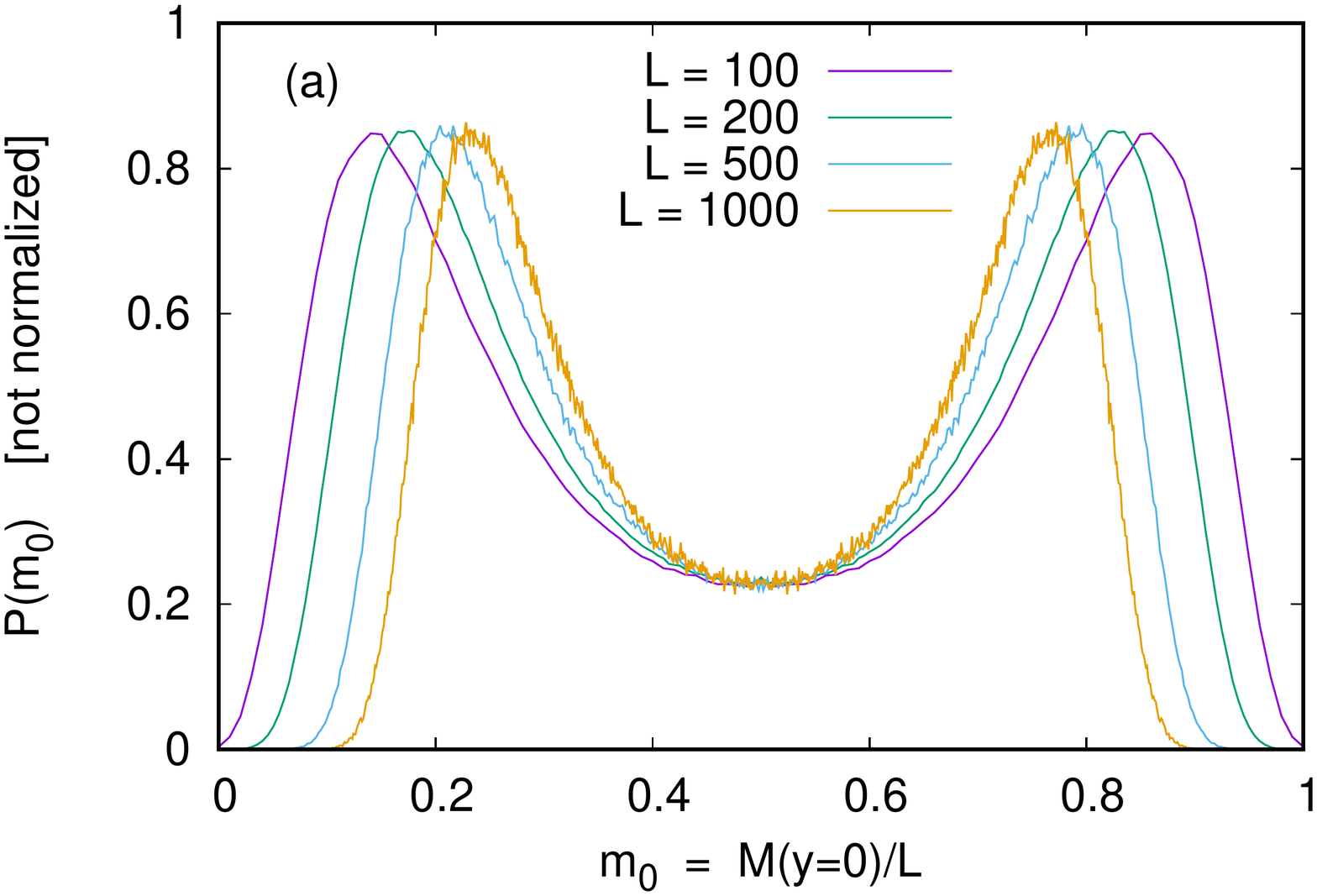}
\vglue -1.0cm
\includegraphics[scale=0.30]{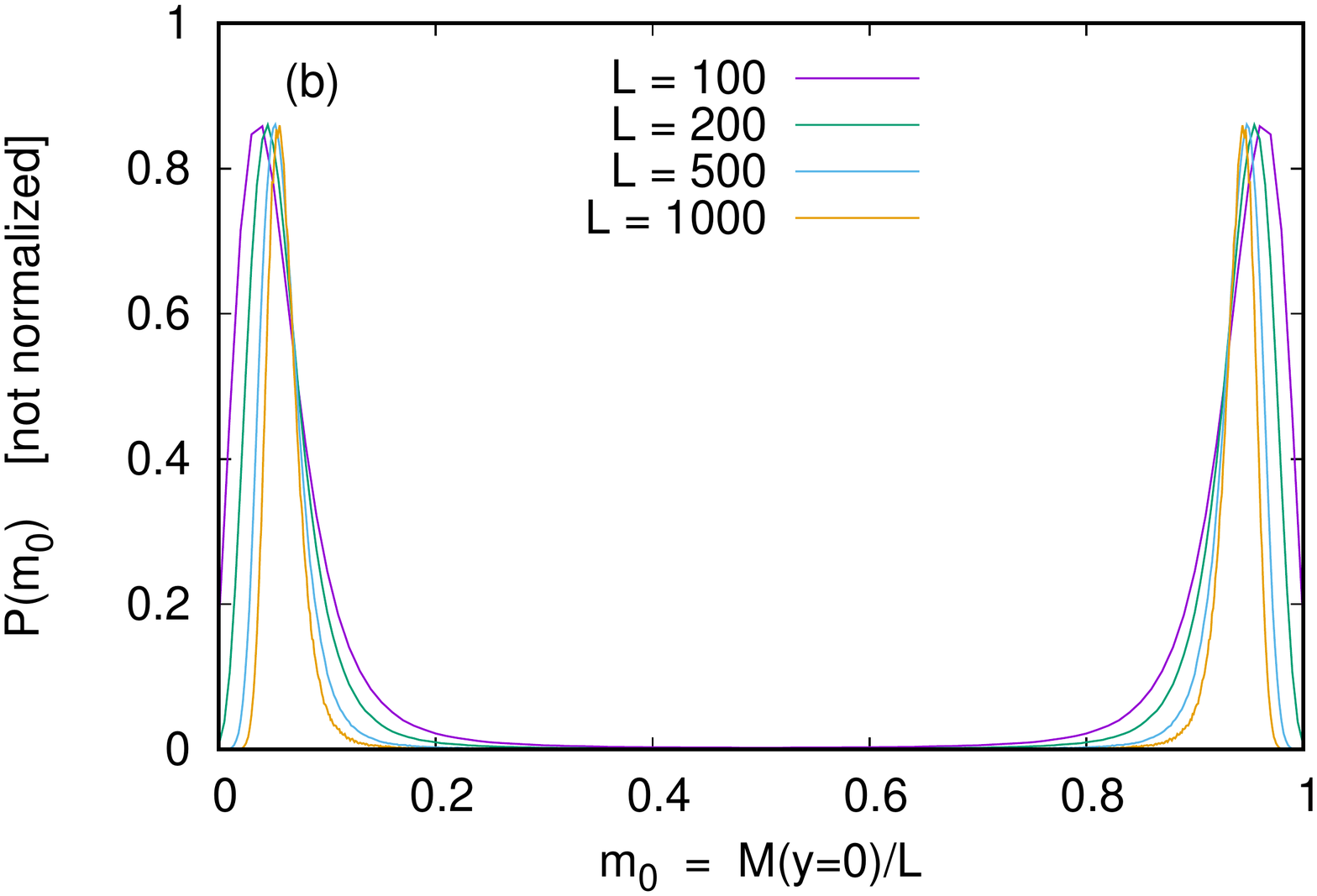}
\par\end{centering}
\caption{\label{figr4} (color online) Non-normalized distributions of the Ising magnetization densities 
  at $y=0$ for $k=2$ (panel (a)) and for $k=3$ (panel (b)).}
\end{figure}

These simulations also show that the spatial extents of walks in the $x$ and $y$ directions scale with the 
Flory exponent for $k=1$ and $k=2$, while they are very different for $k>2$. More precisely, the longitudinal
r.m.s. end-to-end distance $\sqrt{\langle x^2\rangle} $ increases for large $N$ linearly with $N$, while 
$\sqrt{\langle y^2\rangle}$ stays finite and reaches a constant (see Fig.~2). Indeed, we find also that distributions of $y$
do not follow power laws for $k>2$ (as they do for $k=1$ and 2), but are exponentials (data not shown). All 
this means that self avoiding walks are strongly attracted to the spine for $k \geq 3$, to the extent 
that the transverse correlation length is finite, and the growth is ballistic in the direction of the spine. 
Thus all aspects of criticality of the model on single planes are gone.

{\it Ising model}: We simulated the Ising model with the Wolff algorithm \cite{Wolff}. Boundary conditions were 
periodic at the sides $x=0$ and $x=L$, but open at $y=L$. We only show data for the bulk critical temperature, 
as we saw no hint that the critical point is shifted as it happened for SWA's. The main observables were the 
average squared magnetization at fixed $y$ and the finite size scaling of its distribution at $y=0$. Plots of 
$\langle m^2(y)\rangle$ for $L=1000$ are shown in Fig.~3. All curves fall off for large $L$ because of the open 
boundary conditions at $y=L$. For small $y$ we see that the curve is flat for $k=2$, as we expect. It scales 
for $k\geq 3$ as 
\be
   \langle m^2(y)\rangle \sim y^{-2\beta/\nu}
\ee
with $\beta/\nu = 1/8$ as expected for a normal surface with nonzero applied surface field. This indicates 
that the mutual reinforcement of the $k$ sheets is sufficient to order the system at the spine.

The same is also indicated by the distributions of the magnetization at $y=0$. We show in Fig.~4 results
for $k=2$ (top panel) and $k=3$ (bottom panel. All distributions are not normalized. For $k=2$ we see the 
usual finite size scaling for second order phase transitions \cite{Wilding}
\be
    P(m) \propto \Phi(m L^{\beta/\nu}),
\ee
i.e. the curves are just rescaled when $L$ is changed. This is not the case for $k=3$, where the two peaks
become sharper as $L$ increased, and the valley between them becomes deeper. This is the hall mark of a 
first order phase transition. Moreover, due to the increasing depth of the valley, magnetization 
switches become more and more rare with increasing $L$. Thus ergodicity is for 
large systems already broken at $T=T_c$, and not only for $T<T_c$ as for single-sheeted lattices.
Notice however that at the same time we have power behaved correlations at large $y$, i.e. we have 
a hybrid case where the discontinuous jump of the order parameter at $y=0$ coexists with a continuous 
transition for $y\gg 0$.

\begin{figure}
\begin{centering}
\includegraphics[scale=0.30]{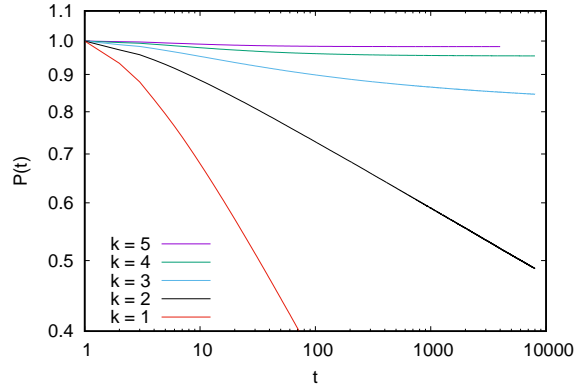}
\par\end{centering}
\caption{\label{figr5} (color online) Probabilities $P(t)$ that a percolation cluster at $p=p_c$ continues 
   to grow for time $\geq t$. Clusters start growing at the line $y=0$.}
\end{figure}

{\it Percolation}: Finally we shall discuss the case of percolation. More precisely, we study site 
percolation by means of the Leith algorithm, which allows us to start cluster growth from a single site 
at a specific value of $y$. We concentrate of course on clusters grown from $y=0$. We use the breadth-first 
version of the Leith algorithm, and denote by $t$ the `time', i.e., the number of spreading steps needed
to reach a site.

In a first set of runs we used $p=p_c = 0.59274605$ \cite{Jacobsen}, stopped the growth at $t=t_{\rm max}$, 
and used lattices that were large enough so that the boundaries were never reached. 
In Fig.~5 we show the probabilities $P(t)$ that a cluster starting at $y=0$ continues to grow at least 
for $t$ time steps. For $k=1$ and $k=2$ it is known that
$P(t)$ obeys a power law $P_k(t) \sim t^{-\delta_k}$ with $\delta_k = \beta_k/\nu/d_{\rm min}$ with
$\nu = 4/3$ \cite{Nijs}, $d_{\rm min} = 1.13077(2)$ \cite{Zhou}, $\beta_1 = 4/9$, and $\beta_2 = \beta = 5/36$
\cite{Vanderzande}. For $k\geq 3$ it seems that $P(t)$ converges towards a positive value for
$t\to\infty$. Assuming an ansatz $P(t) = P(\infty) - a/t^\Delta$, we obtained $P(\infty) = 0.813(5)$ and 
$\Delta = 0.22(2)$ for $k=3$, while we got $P(\infty) = 0.954(4)$ and $\Delta = 0.6(1)$ for $k=4$.

Notice that this positivity of $P(\infty)$ is not, in contrast to the case of SAW's, due to the fact that
the critical point has shifted for $k>2$, as we shall see explicitly below.
Since $P(\infty)$ is also the probability that a randomly chosen site is located on the infinite cluster
and is considered as the order parameter, we thus conclude that the order parameter is finite at the 
critical point for $k\geq 3$ and for $y=0$, i.e. the model shows some aspect of a first order transition.
To further support this, we made a second set of runs -- also at $p=p_c$ -- on lattices of size $L_x 
\times L_y$ with $L_y \gg t_{\rm max}^{1/d_{\rm min}} \gg L_x$ and with periodic b.c. in the $x$-direction.
Also here we started the growth at $y=0$. But now times were so large that clusters that survived for long 
times covered the $x$-axis $\{0 < x < L_x,\; y=0\}$ uniformly. If the above scenario with  $P(\infty) > 0$ is 
correct, we expect then for each run essentially two possibilities: Either the cluster dies soon and the 
density $\rho_0$ of `wetted' (or `infected') sites on the $x$-axis is small ($O(1/L_x)$), or this density 
is finite and turns for $L_x\to\infty$ towards $P(\infty)$. 

\begin{figure}
\begin{centering}
\includegraphics[scale=0.30]{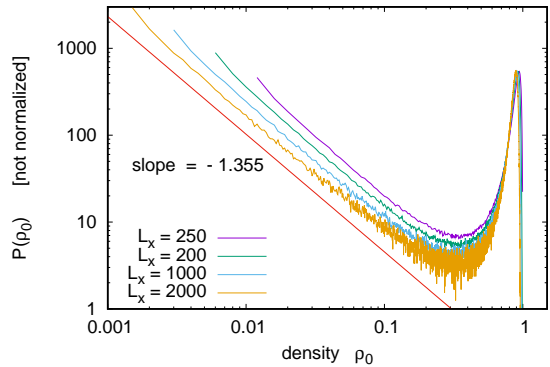}
\par\end{centering}
\caption{\label{figr6} (color online) Non-normalized distributions of the density $\rho_0)$ of sites on 
   the axis $y=0$ ``wetted" by clusters that start to grow at $y=0$. In these simulations $L_x$ was 
   finite, but $L_y$ was so large as to be effectively infinite. The left hand peak corresponds to clusters
   that died early, while the r.h.s peak corresponds to clusters that grow to values $y\gg L_x$.}
\end{figure}

Data obtained for $k=3$ are shown in Fig.~6 and confirm this perfectly: While there are two broad peaks of 
the density distribution $P(\rho_0)$ for finite $L_x$, they tend towards delta-peaks for $L_x\to\infty$. 
We show non-normalized distributions where the right hand peaks have the same heights, because then it is
most easy to verify that the peak positions do not move with $L$, i.e. the density is finite and not fractal. 
At the left hand peak we see a scaling law $P(\rho_0) \sim \rho_0^{-\sigma}$ with $\sigma = 1.355(10)$,
indicating that there is a new non-trivial exponent for small clusters starting to grow at $y=0$.

\begin{figure}
\begin{centering}
\includegraphics[scale=0.30]{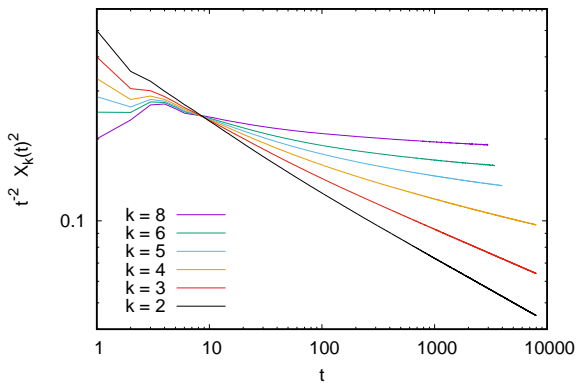}
\par\end{centering}
\caption{\label{figr7} (color online) Log-log plots of the (squared) cluster growth in $x$-direction,
   for $k=2,3,\ldots 8$. For $k=2$ we have of course the usual growth of 2-d percolation clusters,
   but for $k>2$ the growth follows power laws with larger exponents.}
\end{figure}

Clusters grown at $y=0$ for $k\geq 3$ are strongly anisotropic, as seen also by the growth of 
$X_k(t)^2 = \langle x^2\rangle$ and $Y_k(t)^2 = \langle y^2\rangle$. Here the seed is assumed to be at
$x=y=0$, and the averages are over all active sites at time $t$. Data for $t^{-2}X_k(t)^2$ obtained during 
the first set of runs are shown in Fig.~7. We do not show the data for $Y_k(t)^2$ since they are trivial: 
For all $k$, we have $Y_k(t)^2 \sim t^{2/d_{\rm min}}$. This is not so for $X_k(t)^2$ with $k>2$. Figure 
7 strongly suggests non-trivial power laws $X_k(t)^2 \sim t^{z_k}$ for $k=3$ and $k=4$ (with $z_3 = 1.821(1)$ 
and $z_4 = 1.91(1)$). But for $k \geq 5$ we no longer are able to distinguish between such non-trivial 
scaling laws and trivial power laws $X_k(t)^2 \sim t^2$ with very long transients.

\begin{figure}
\begin{centering}
\includegraphics[scale=0.30]{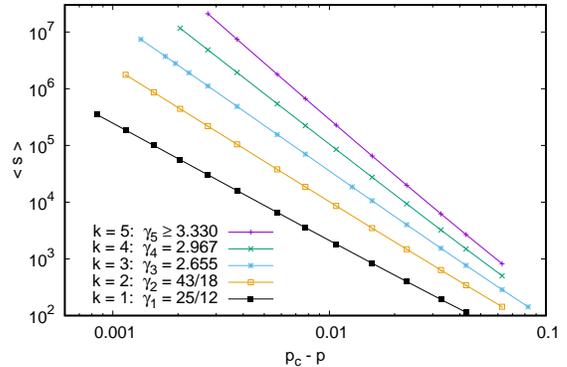}
\par\end{centering}
\caption{\label{figr8} (color online) Log-log plots of average cluster sizes, for subcritical clusters 
   grown from seeds at $y=0$ on infinitely large lattices. For $k\leq 2$ we recover the known scaling
   laws, for $k\geq 3$ we find again scaling laws with new exponents.}
\end{figure}

Finally, in a last set of runs, we estimated average cluster sizes $s$ for subcritical values of $p$. This 
time both $L_x$ and $L_y$ were large enough so that all clusters died before boundaries were reached.
All clusters started to grow at $y=0$. In ordinary percolation \cite{Stauffer} we have 
$s \sim (p_c - p)^{-\gamma}$, where $\gamma=43/18$ for the bulk and $\gamma=\gamma_1=25/12$ for clusters
growing at a boundary \cite{Vanderzande}. From Fig.~8 we see that this generalizes to arbitrary $k\geq 1$,
with nontrivial exponents $\gamma_k$ at least for $k \leq 5$. For large $k$ we see however large corrections
to scaling, due to which we cannot give reliable estimates for $\gamma_k$ with $k\geq 5$.
In any case, these simulations show clearly that the critical point is not shifted for $k>2$.

{\it Discussion:} One reason for this study was the aim of finding and studying novel cases of hybrid phase 
transitions. These are transitions that show features of second order transitions like anomalous scaling laws, but 
where also order parameters jump discontinuously as in first order transitions. During recent years many such 
cases were found \cite{Aizenman,Causo,Toninelli,Schwarz,Goltsev,Baxter,Bizhani,Buldyrev,Son,Bar,Boettcher,Cai,Grassberger},
but they are still not well understood.
In two of our three models (Ising, percolation) the transition is indeed hybrid, while it is not for SAW's.
The fact that the transition is hybrid in the Ising case and for directed percolation has also been found 
in \cite{Igloi:1993,Igloi:1993a,Igloi:2016}.

More fundamentally, it is of interest to study what happens to phase transitions
when the underlying structure is a regular lattice {\it nearly} everywhere -- but
not everywhere -- and has non-trivial global topology. The main examples for this are regular
lattices with boundaries. To my knowledge,
polymer networks \cite{Duplantier} are the only well studied model where systems with
complex topologies are formed by glueing together simple systems. There the situation is
similar to the percolation case studied here, in the sense that some critical exponents
depend on the topology while others do not. We should point out that lattice defects alone -- even
extended ones as in \cite{brochette}, where one line in a 2-d lattice is different from the rest --
do not give rise in the Ising model and in percolation to phenomena similar to the ones discussed here.

Finally let us discuss possible experimental realizations. The prime example of a 3-booklet is of course 
three three soap films meeting at 120 degrees. Self avoiding walks can be studied in this geometry by
placing a long polymer (e.g. DNA) close to the line where they meet.
Due to the changed connective constant, the polymer will be
drawn towards it. When it is there, it will be very much stretched to stay as close to the line as 
possible. The problem is of course more complicated than our toy model, because the soap films are 
elastic and will deform in response to the polymer \cite{Mathe}. Thus a detailed quantitative 
analysis would be not easy. Nevertheless, the basic predictions of our model should apply, and it would be 
an interesting question how much they are modified by film elasticity. 

Instead of soap films one could of course also consider lipid bilayers or other biomembranes, although 
it might not be so easy to make clean 3-booklets. The advantage over soap films would then be that 
biomembranes show various phase transitions \cite{Nagle,Mohwald,Blume,Guidi}. Studying these transitions on 
a booklet geometry could then open a wide range of novel phenomena.

I thank Michel Pleimling for enlightening correspondence. I am also indebted to Ferenc Igl\'oi for 
pointing out Refs. \cite{Igloi:1993,Igloi:1993a} and for sending me \cite{Igloi:2016} prior to publication.

\end{document}